\documentclass[a4paper,11pt]{article}

\pdfoutput=1 

\usepackage{jinstpub} 
                    
\title{Forward scattering effects on muon imaging}
\author[a,1]{H. G\'{o}mez,\note{Corresponding author.}}
\author[b,c]{D. Gibert,} 
\author[d]{C. Goy,}  
\author[e]{K. Jourde,} 
\author[d]{Y. Karyotakis,} 
\author[a]{S. Katsanevas,} 
\author[f]{J. Marteau,} 
\author[g]{M. Rosas-Carbajal,} 
\author[a]{A. Tonazzo.} 

\affiliation[a]{AstroParticule et Cosmologie, Universit\'{e} Paris Diderot, CNRS/IN2P3, CEA/IRFU, Observatoire de Paris, Sorbonne Paris Cit\'{e}, 75205 Paris Cedex 13, France}
\affiliation[b]{OSUR - G\'eosciences Rennes (CNRS UMR 6118), Universit\'e Rennes 1, Rennes, France.}
\affiliation[c]{National Volcano Observatories Service, Institut de Physique du Globe de Paris (CNRS UMR 7154), Paris, France.}
\affiliation[d]{Laboratoire d'Annecy de Physique des Particules (LAPP), Universit\'e de Savoie, CNRS/IN2P3, Annecy, France}
\affiliation[e]{Bureau de Recherches G\'{e}ologiques et Mini\`{e}res (BRGM), Orl\'eans, France.}
\affiliation[f]{Institut de Physique Nucl\'{e}aire de Lyon (IPNL) - Universit\'{e} de Lyon UCBL, Lyon, France}
\affiliation[g]{Institut de Physique du Globe de Paris, Sorbonne  Paris Cit\'e, Univ Paris Diderot, UMR 7154 CNRS, Paris, France.}

\emailAdd{hgomez@apc.univ-paris7.fr}

\abstract{Muon imaging is one of the most promising non-invasive techniques for density structure scanning, specially for large objects reaching the kilometre scale. It has already interesting applications in different fields like geophysics or nuclear safety and has been proposed for some others like engineering or archaeology. One of the approaches of this technique is based on the well-known radiography principle, by reconstructing the incident direction of the detected muons after crossing the studied objects. In this case, muons detected after a previous forward scattering on the object surface represent an irreducible background noise, leading to a bias on the measurement and consequently on the reconstruction of the object mean density. Therefore, a prior characterization of this effect represents valuable information to conveniently correct the obtained results. Although the muon scattering process has been already theoretically described, a general study of this process has been carried out based on Monte Carlo simulations, resulting in a versatile tool to evaluate this effect for different object geometries and compositions. As an example, these simulations have been used to evaluate the impact of forward scattered muons on two different applications of muon imaging: archaeology and volcanology, revealing a significant impact on the latter case. The general way in which all the tools used have been developed can allow to make equivalent studies in the future for other muon imaging applications following the same procedure.}

\keywords{Models and simulations, Simulations methods and programs}

\arxivnumber{1709.05106} 

\begin{document}
\maketitle
\flushbottom

\section{Introduction}
\label{sec:Introduction}

The idea to use muons produced in the Earth's atmosphere by cosmic rays as a scanning method of anthropic or geological structures, the so-called muon imaging, was proposed soon after the discovery of these muons \cite{Anderson1, Anderson2, Auger}. Muon imaging leverages the capability of cosmic muons to pass through hundreds of metres or even kilometres of ordinary matter with an attenuation mainly related to the length and density of this matter encountered by the muons along their trajectory before their detection \cite{Nagamine}. As this attenuation is principally caused by muons absorption and scattering, muon imaging can be mainly performed using two different techniques. The first one is the so-called transmission and absorption muography \cite{Lesparre_2010, Marteau_2012}. This technique relies on the well-known radiography concept (widely used, for example, in medicine with X-rays), based on the muon energy loss and its consequently probability to cross a given amount of material. The second is known as deviation muography, which relies on the measurement of the muon track deviation to determine the object density \cite{Borozdin, Procureur}. For the first technique, studying all the directions for which muons go through the studied object, and knowing its external shape, it is possible to obtain a 2D mean density image. Thus, muon imaging provides a non-invasive and remote scanning technique utilisable even for large objects, where the detection set-up may be relatively far away from the -- potentially dangerous -- target (e.g. domes of active volcanoes or damaged nuclear reactors).

One of the first studies performed based on muon imaging dates from 1955, being the scanning of the rock overburden over a tunnel in Australia
\cite{George}. Later, other applications from mining \cite{Malmqvist} to archaeology were proposed, both in the 70s. For the latter case, some measurements have been already performed, as the exploration of the Egyptian Chephren \cite{Alvarez} and Khufu \cite{SPNature} pyramids. Nowadays, thanks to the improvements on the detector performance, and also to their autonomy and portability, muon imaging reveals itself to be a scanning technique competitive and complementary to others non-invasive methods as seismic and electrical resistivity tomography or gravimetry. This has led to its proposal and utilisation in a wide range of fields.

In addition to the above-mentioned applications (archaeology and mining), two others stand out. The first one is related with geophysics, more precisely with the monitoring of volcanoes. This has an important benefit both from a scientific and social point of view. The continuous monitoring of volcanoes helps to understand their internal dynamics, a key feature in the risk assessment. The other application, more related with particle physics, was motivated by the necessity to characterize the overburden of underground laboratories hosting various experiment detectors. It is worth mentioning other applications related to civil engineering and nuclear safety. For the first one, it will be possible for example to scan structures looking for defects. For the case of nuclear safety, set-ups looking for the transport of radioactive materials and wastes already work in cooperation with homeland security agencies. Moreover, the study of nuclear reactors looking for structural damages have been already used, as in the case of the recent Fukushima nuclear power plant accident \cite{Kume}, and it is being considered as a remote scanning method.

As mentioned, the improvement on the detectors used for muon imaging has been one of the main reasons for the renewal of this technique. Better detectors provide a better angular resolution for the muon direction reconstruction and improve the precision of the density radiography. Nonetheless, the background muon flux rejection remains a key procedure for the structural imaging with muons. An important potential noise source, specially in the measurements based on the transmission and absorption muography, is the forward scattering of low energy muons on the object surface, reaching afterwards the detector. This effect mimics through-going particles since the reconstructed direction of the scattered particle points towards the target. The result is an increase of the total number of detected particles, as if the target's opacity was lower than its actual value, leading to a systematic underestimation of the density \cite{nishiyama,3d_souf}. Being produced by muons, these events can not be rejected by particle identification techniques, representing an irreducible background. For this reason, an evaluation of the magnitude of this effect is mandatory to conveniently correct the reconstructed object density.

In this work, a general evaluation of forward scattering of muons has been performed by Monte Carlo simulations. The aim was to develop a versatile tool to be able to evaluate this process for different object geometries and compositions, due to the increasing number of proposed applications based on muon tomography. The main features and results are presented in section \ref{sec:Scattering}. Then the impact of this process on the muon imaging capability has been evaluated defining a signal to background parameter. Two physics cases have been studied in section \ref{sec:SBratio}. The first one concerns an archaeological target, the Apollonia tumulus near Thessaloniki in Greece, and the second one \textit{La Soufri\`ere} volcano in Guadeloupe Islands of the Lesser Antilles. Finally, a summary of the different results and the main conclusions extracted from them are compiled in section \ref{sec:conclusions}.

\section{Evaluation of the forward scattering of muons}
\label{sec:Scattering}

As mentioned in the introduction, low-energy cosmic muons can change their original direction after interacting with the target or any other object in the surroundings before their detection. As muon imaging is based on the reconstruction of the detected muons direction, these muons would forge the measurement. As a consequence, the determination of the target's internal structure and the corresponding reconstructed mean density will be affected. 

Muons trajectory deviation is mainly driven by their interaction with matter via multiple Coulomb scattering. The resulting deflection angular distribution, theoretically described by the Moli\`{e}re theory \cite{Bethe}, roughly follows a Gaussian,

\begin{equation}
\frac{dN}{d\alpha} = \frac{1}{\sqrt{2\pi}\alpha_{MS}} e^{-\frac{\alpha^2}{2\alpha^{2}_{MS}}}
\label{eq:MoliereGauss}
\end{equation}

\noindent which is centred in zero (i.e. no deflection happens), having a standard deviation $\alpha_{MS}$:

\begin{equation}
\alpha_{MS}=\frac{13.6 MeV}{\beta cp}Q\sqrt{\frac{x}{X_0}}\left (1 + 0.038 ln(x/X_0) \right )
\label{eq:MoliereAngle}
\end{equation}

\noindent where $\beta$ is the relativistic factor, $p$ the muon momentum in MeV/$c$, $x$ the material thickness and $Q$ the absolute electric charge of the muon. $\alpha_{MS}$ also depends on the radiation length ($X_0$) which is empirically given by

\begin{equation}
X_0 \approx \frac{716.4 g/cm^2}{\rho} \frac{A}{Z(Z+1)log(287/\sqrt{(Z)})}
\label{eq:RadialtionL}
\end{equation}

\noindent with $Z$ and $A$ the atomic and mass numbers respectively and $\rho$ the material density. This reveals the relationship of the multiple Coulomb scattering with the properties of the studied material.

\begin{figure}[ht]
\begin{center}
\includegraphics[width=0.5\textwidth]{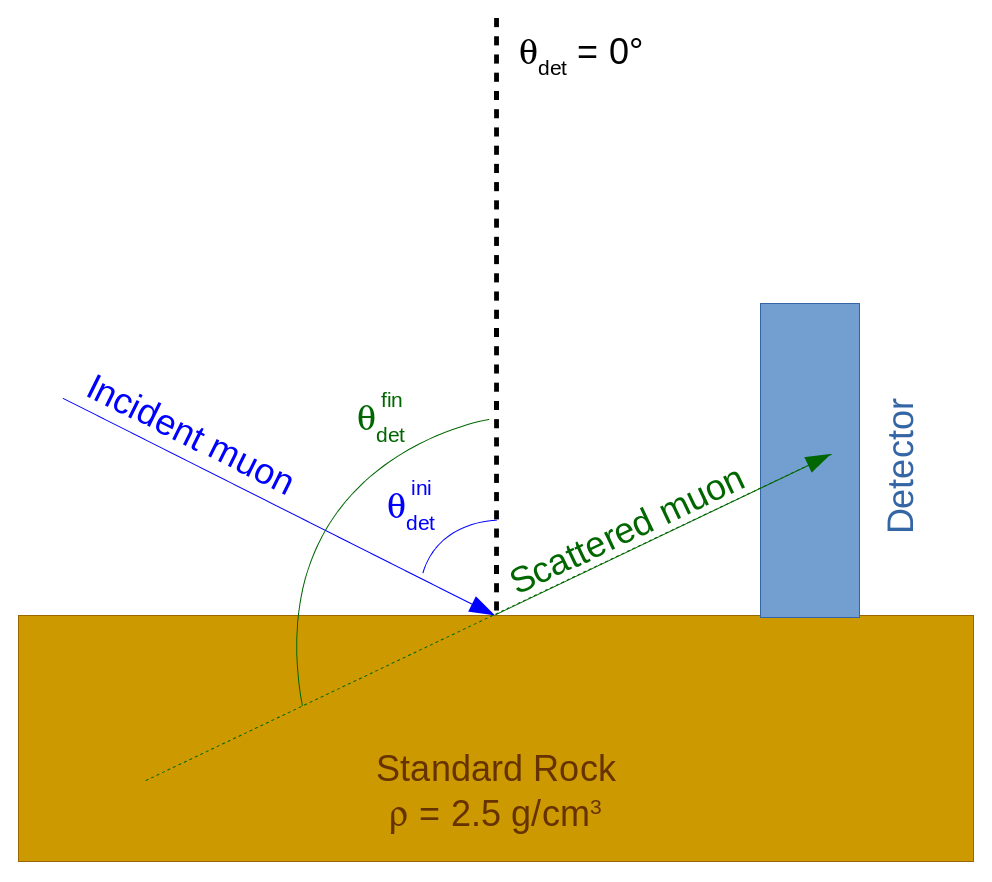}
\caption{Schema of the defined geometry to perform the general studies of forward scattering of muons.}
\label{General_Diff_Schema_Img}
\end{center}
\end{figure}

Different works (see for example \cite{Schneider}) provide analytical solutions to the angular distribution of deflected muons after traversing an object with a determined geometry and composition. Besides, other relevant features, as the higher scattering probability for lower energy muons, are also demonstrated in these studies. However, the increasing number of different applications proposed for muon tomography, implies a large variety of objects dimensions, shapes and compositions, being less evident to obtain an analytical estimation of the forward scattering process suitable for all these cases. In this context, Monte Carlo simulations represent a useful tool for the study of muon scattering process, versatile enough to adapt them to the main features of each particular case. As first step on the development of these simulations, a general evaluation of the muons forward scattering has been carried using the Geant4 simulation tool-kit \cite{Geant4}. It allows the simulation of the 3D muon transport through the defined geometry taking into account the energy loss and trajectory variations due to multiple Coulomb scattering as well as to ionization, bremsstrahlung, pair production and multiple inelastic scattering. Considering these possible processes, results can be compared with the estimations given by the analytical formulas above-mentioned. A scheme of the simulated set-up is shown in figure \ref{General_Diff_Schema_Img}. For this case, generated muons are thrown to a fixed point on a standard rock surface (with a density of 2.5 g/cm$^{3}$). In the case of scattered muons, the direction changes, in zenith and/or azimuth angles, can be evaluated.

\begin{figure}[ht]
\begin{center}
\includegraphics[width=0.45\textwidth]{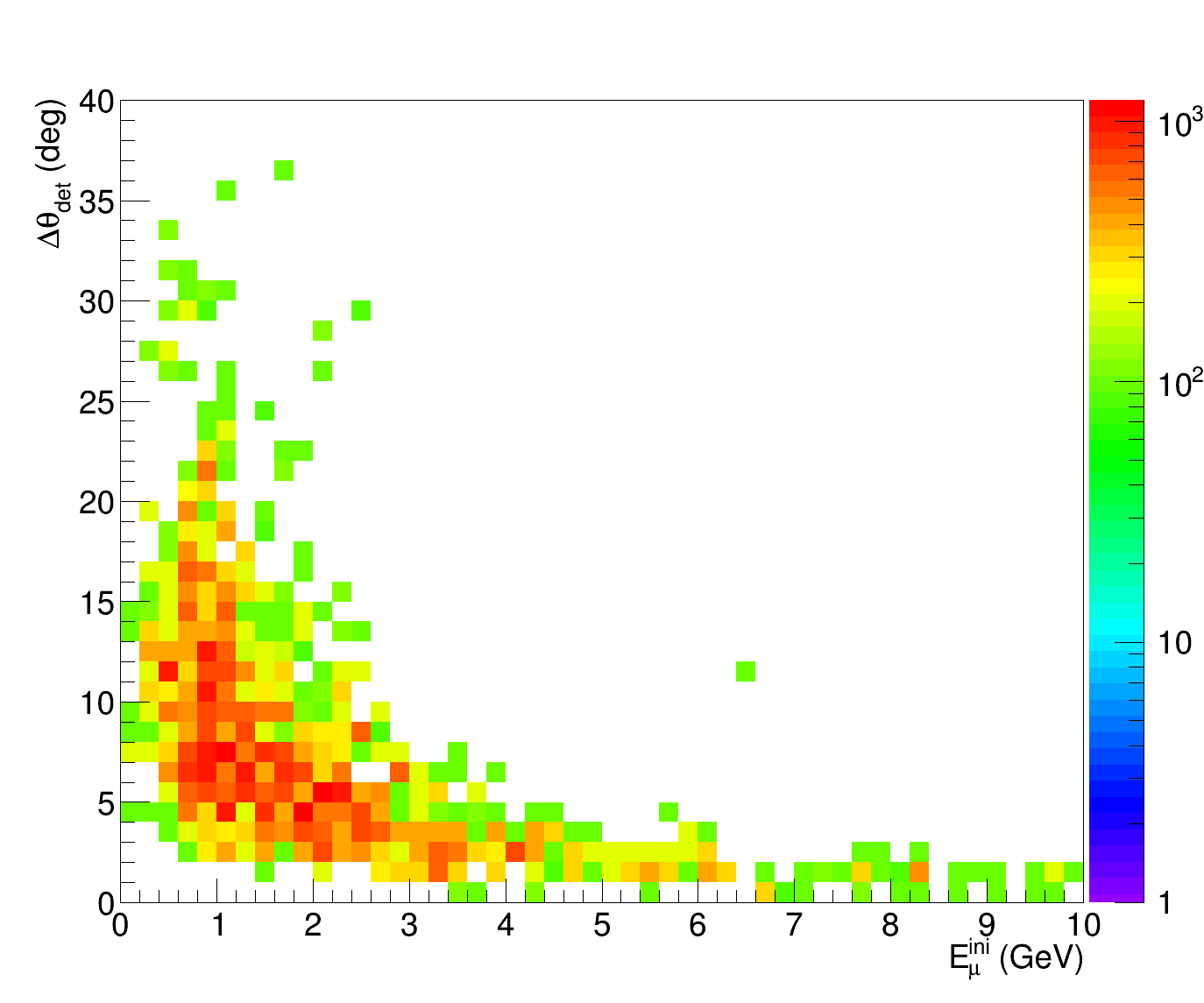}
\includegraphics[width=0.45\textwidth]{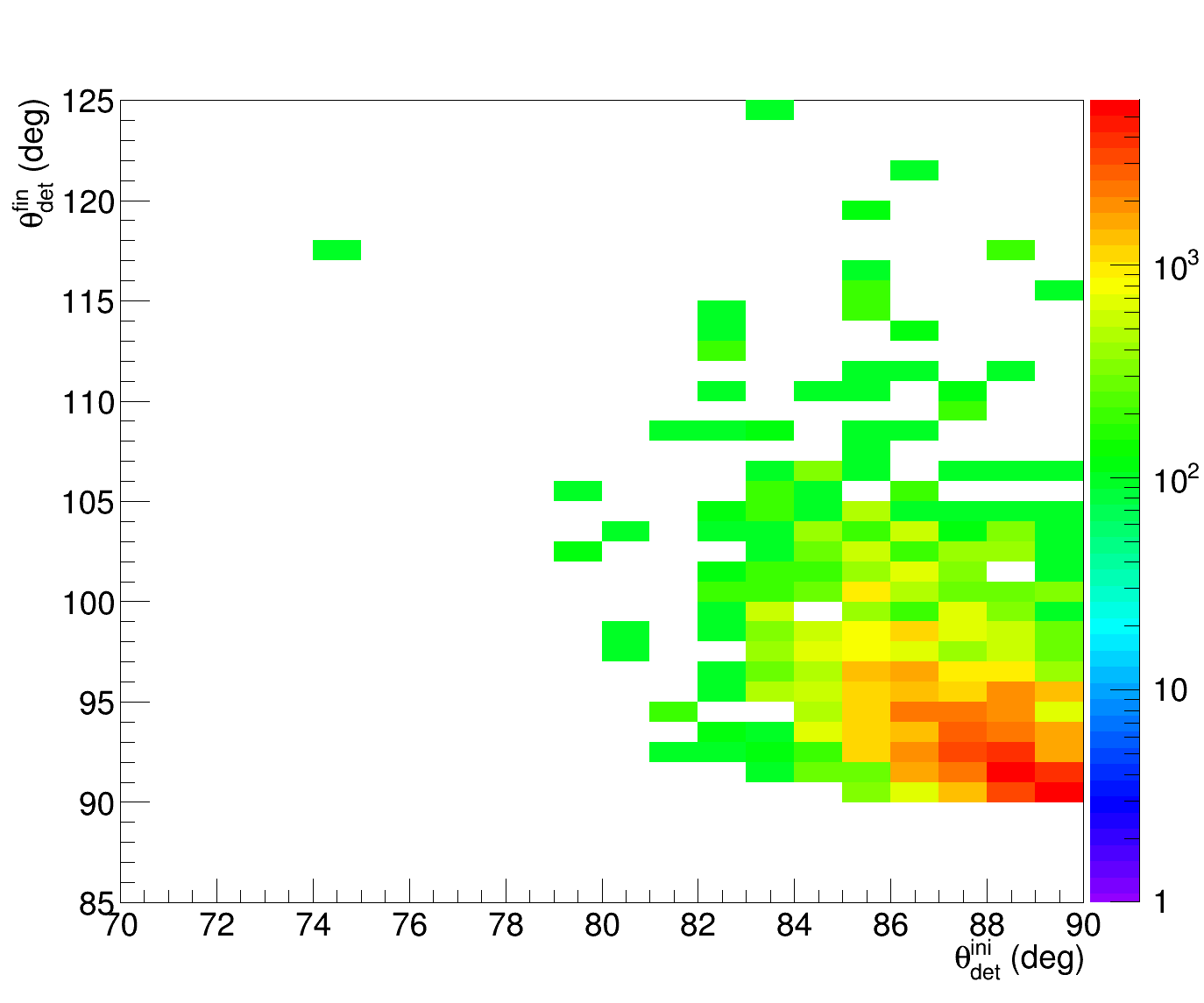}
\includegraphics[width=0.45\textwidth]{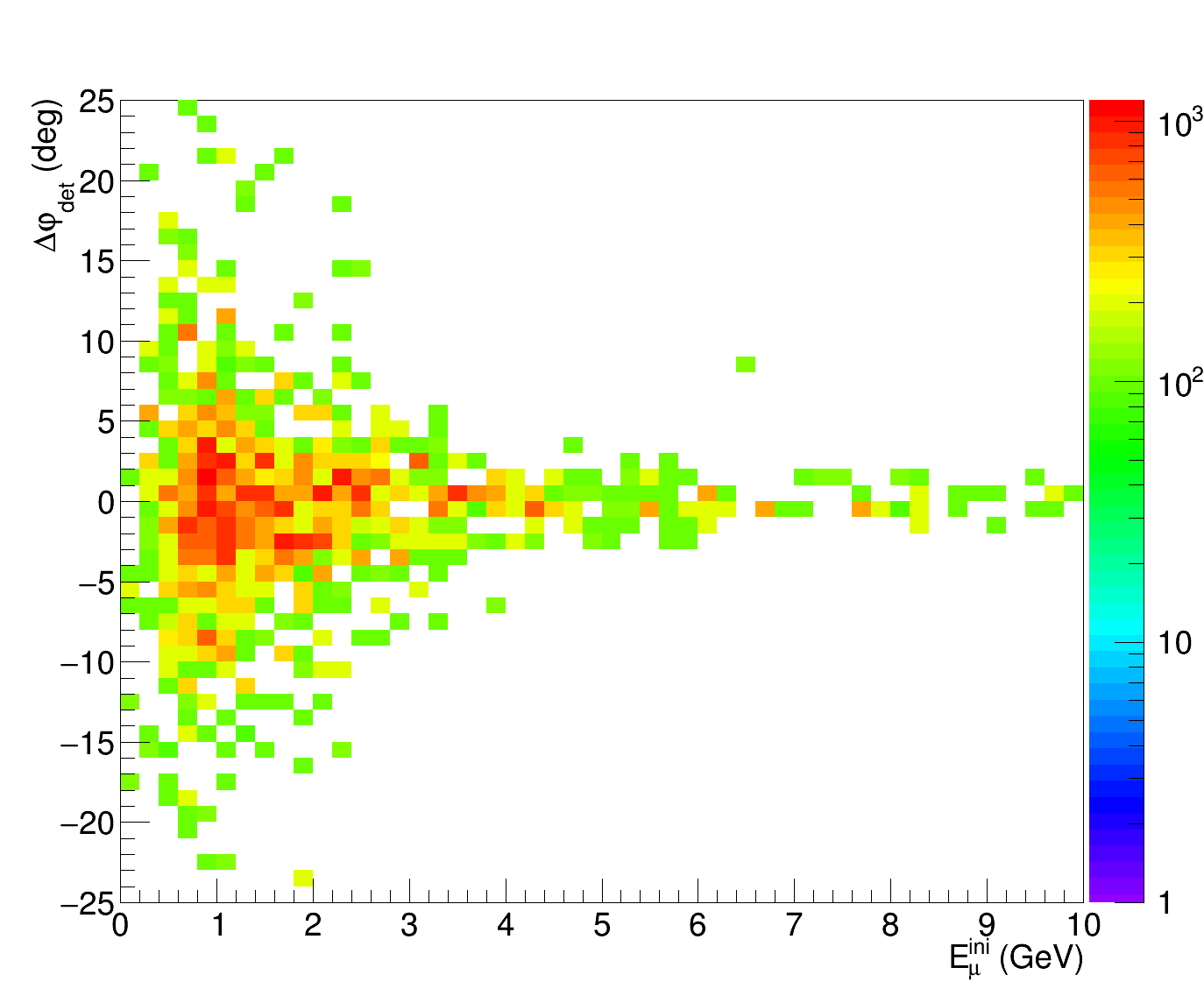}
\includegraphics[width=0.45\textwidth]{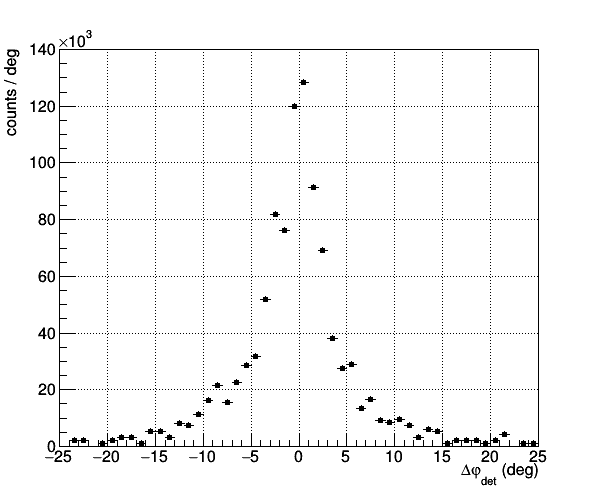}
\caption{Summary plots of the results of the general study of forward scattering of muons (see text for details about the study). Top left: Difference on the zenith angle ($\Delta\theta_{det}$ = $\theta^{fin}_{det}$ - $\theta^{ini}_{det}$) with respect to the initial muon energy ($E_{\mu}^{ini}$). Top right: Correlation between the initial and final zenith angles ($\theta^{ini}_{det}$ vs. $\theta^{fin}_{det}$) for all the muon energies considered. Bottom left: Difference on the azimuth angle ($\Delta\varphi_{det}$ = $\varphi^{fin}_{det}$ - $\varphi^{ini}_{det}$) with respect to the initial muon energy ($E_{\mu}^{ini}$). Bottom right: $\Delta\varphi_{det}$ distribution for all the muon energies considered. A Gaussian distribution as predicted by Moli\`{e}re theory (Equations \ref{eq:MoliereGauss} - \ref{eq:RadialtionL}) is observed.}
\label{General_Diff_Img}
\end{center}
\end{figure}

A first set of simulations were performed in order to evaluate the general features of the muon forward scattering. In the previously described set-up, muons up to 10 GeV, with a zenith incident angle ($\theta^{ini}_{det}$) between 70$^{\circ}$ and 90$^{\circ}$ and an azimuth incident angle $\varphi^{ini}_{det}$ = 0$^{\circ}$ were generated. It is worth mentioning that by the set-up definition $\theta^{ini}_{det}$ = 0$^{\circ}$ implies muons perpendicular to the rock surface, while $\theta^{ini}_{det}$ = 90$^{\circ}$ corresponds to tangential ones. Figure \ref{General_Diff_Img} summarizes the results of this general simulation, leading to some conclusions about the muon forward scattering studied in these simulations. First, it is observed that this process is negligible if the muon energy is higher than 5 GeV, independently of the incident direction. For the lower energy muons, most of the ``efficient'' scattering processes (i.e. when the scattered particle exits the medium) occur if $\theta^{ini}_{det}$ is higher than 85$^{\circ}$ and do not exist if $\theta^{ini}_{det}$ is lower than 80$^{\circ}$. That means that only low energy muons with incident directions close to the surface tangent are likely to be scattered on the object surface and to induce a signal in the detector. For these muons the angular deviation can reach up to 25$^{\circ}$ both for the zenith and azimuth angles. By the simulation set-up definition, only the azimuth scattering angle ($\Delta\varphi_{det}$) has been registered for the whole angular range. As presented in figure \ref{General_Diff_Img}, the $\Delta\varphi_{det}$ distribution for all the muon energies considered is in agreement with the Gaussian predicted by Moli\`{e}re theory, as well as the other extracted conclusions agree with the analytical predictions \cite{PDG}.

\begin{figure}[]
\begin{center}
\includegraphics[width=0.98\textwidth]{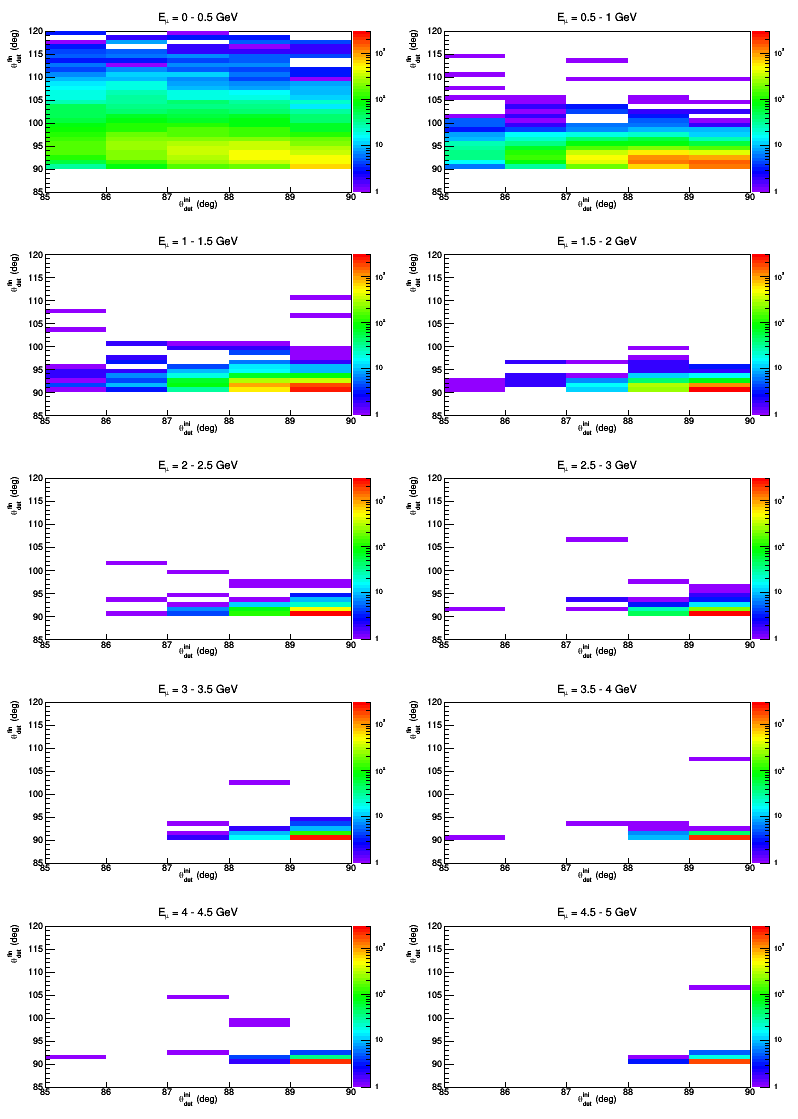}
\caption{Correlation between the initial and final zenith angles ($\theta^{ini}_{det}$ vs. $\theta^{fin}_{det}$) from the general study of forward scattering of muons (see text for details about the study). The correlation plots are divided in 0.5 GeV windows between 0 and 5 GeV for the initial simulated muon energy ($E_{\mu}^{ini}$). These plots, correspond to 10$^{8}$ simulated muons with incident angles between 85$^{\circ}$ and 90$^{\circ}$ and energies between 0 and 5 GeV (both of them homogeneously distributed). They are used as PDF for further estimations on the forward scattered muon flux.}
\label{Diff_PDF_Ewindows_Img}
\end{center}
\end{figure}

Taking into account this general information and having checked the agreement between the general simulations and the analytical predictions, a more detailed simulation, optimizing the initial muon sampling was performed. The objective was to establish a probability density function (PDF) to further estimate the background due to forward scattered muons that could be detected during a muon imaging measurement and should be considered in the image analysis. With this aim 10$^{8}$ muons homogeneously distributed up to 5 GeV and with $\theta^{ini}_{det}$ between 85$^{\circ}$ and 90$^{\circ}$ (all with $\varphi^{ini}_{det}$ = 0$^{\circ}$) were generated and simulated in the described Geant4 framework. The generated PDF provides a probability value $\mathcal{P}$($\theta^{ini}_{det}$, $\theta^{fin}_{det}$, E$^{ini}_{\mu}$) depending on the initial and final zenith angle and the initial muon energy. A summary plot of the generated PDF divided in 0.5 GeV energy windows is shown in figure \ref{Diff_PDF_Ewindows_Img}. 

At this point it is worth mentioning that for the studies presented in this work (summarized in section \ref{sec:SBratio}), the considered composition of the studied objects are the standard rock used to generate the PDF, but also a definition of soil with different composition and density than the rock ($\rho$ = 2.2 g/cm$^{3}$). Moreover, there exist several types of rocks and soils with different compositions and densities typically, between 2.0 and 2.5 g/cm$^{3}$. For this reason the influence of these two parameters in the PDF generation has been evaluated: a set of dedicated simulations have been performed changing the composition and the density of the target to compare their results. The obtained PDFs, including the standard soil case, agree to better than 97 \%. Thus, the PDF presented in figure \ref{Diff_PDF_Ewindows_Img} has been used for all the studies.
 
\section{Signal to background ratio estimations}
\label{sec:SBratio}

The impact of the forward scattered muons in an imaging measurement for a particular object can be evaluated based on simulations as those presented in section \ref{sec:Scattering}. This impact can be expressed as a signal to background ratio (S/B) for a given direction $\theta_z$ - $\varphi_z$. These spherical coordinates correspond to those centred at the detector where $\theta_z$ = 0$^{\circ}$ is the vertical direction and $\varphi_z$ = 0$^{\circ}$ points to the main axis of the studied object. The signal S($\theta_z$, $\varphi_z$) is estimated as not scattered muon flux, so their reconstructed direction corresponds to their initial one. The background B($\theta_z$, $\varphi_z$) represents the scattered muons for which the reconstructed direction points towards the target. 

As mentioned, these evaluations allow the study of a particular object, with its corresponding composition. For this it is necessary to know its external shape, to assume the object mean density (since this is the observable that can be extracted from a muon tomography measurement), and to determine the muon detector position with respect to this object. This allows the estimation of the object length traversed by muons for each direction as well as its surfaces positions with respect to the detector and to the Earth's surface (required for the determination of $\theta_z$ and $\varphi_z$).

For this work two cases have been considered, corresponding to two applications of the muon imaging: the archaeology and the volcanology. For the first one a Macedonian tumulus located near Apollonia (Greece) has been studied \cite{gomez2016arche}. For the second, \emph{La Soufri\`ere} volcano (Guadeloupe island in the Lesser Antilles), already explored by muon imaging, has been taken as reference \cite{jourde2013experimental, jourde_nature_2016}.

\subsection{Archaeology: Apollonia tumulus}
\label{sec:Arche}

As quoted in section \ref{sec:Introduction}, the exploration of archaeological structures is one of the applications for which muon imaging has been proposed since it is non invasive and does not induce any harmful signals (contrary, for example, to vibrations used in seismic tomography). Already suggested in the 60s \cite{Alvarez}, there exist at present different projects based on muon imaging devoted to the study of the internal structure of archaeological constructions (see for example \cite{ScanPyramids, SPNature}). The ARCH\'e project proposes to scan the Apollonia Macedonian tumulus \cite{Arche}. These tumuli are man-made burial structures where the tomb, placed on the ground, is covered by soil, creating a mound which can also contain internal corridors. The geometry and dimensions of these tumuli are variable but they can always be approximated to a truncated cone. In the case of Apollonia tumulus its height is 17 m while the radius of the base and the top are 46 m and 16 m respectively. With this geometry the angle of the slope of the lateral surface of the tumulus is 29.5$^{\circ}$.

In the present study a standard soil composition, with a density of 2.2 g/cm$^{3}$, has been assumed. The detector has been placed 4 m beside the tumulus base (50 m from the tumulus base centre), so muons with zenith angles $\theta_z$ > 63.4$^{\circ}$ are those which will provide information about the structure of the tumulus. With all these features the signal and background, S($\theta_{z}$, $\varphi_{z}$) and B($\theta_{z}$, $\varphi_{z}$) respectively, can be estimated for a given direction $\theta_{z}$ - $\varphi_{z}$. As already described, these coordinates are centred at the detector and $\theta_{z}$ = 0$^{\circ}$ correspond to vertical muons, while $\varphi_{z}$ = 0$^{\circ}$ points towards the centre of the tumulus base.

From the knowledge of the external shape, it is possible to determine the length of tumulus traversed by muons for a given direction L($\theta_{z}$, $\varphi_{z}$) and thus, the corresponding opacity as the product of this length by the density ($\varrho$ = L $\times \rho$). The required minimal muon energy ($E_{min}$) to cross the target of opacity $\varrho$ can be calculated as

\begin{equation}
E_{min} = \frac{a}{b} \times  \left (e^{b \times \varrho} - 1 \right )
\label{eq:Emin}
\end{equation}

\noindent where $a(E)$ and $b(E)$ represent the energy loss coefficients due to ionization and radiative losses respectively. In this case, coefficients corresponding to standard rock summarized in \cite{PDG} have been used, obtaining E$_{min}$ values as a function of $\varrho$. As a cross-check, these E$_{min}$ values have been also estimated from the CSDA range values of standard rock \cite{PDG_muons}. The agreement between both methods is better than 95 \%.

Hence, the expected signal S($\theta_{z}$, $\varphi_{z}$) corresponds to the muon flux on the studied direction with energies higher than E$_{min}$:

\begin{equation}
S (\theta_{z}, \varphi_{z}) = \int_{E = E_{min}}^{\infty} \phi_{\mu}(E, \theta_{z}, \varphi_{z}) dE
\label{eq:Signal}
\end{equation}

To compute the background due to muons forward scattered in the same direction, B($\theta_{z}$, $\varphi_{z}$), two assumptions have been done. First, a point-like detector is considered. This implies that for each scattering point on the tumulus surfaces there is a unique final direction reaching the detector. Second, scattering effects in the azimuth angle are neglected. Since the general muon scattering studies (section \ref{sec:Scattering}) show that these effects are symmetric and mostly below 5$^{\circ}$ for the azimuth angle (see figure \ref{General_Diff_Img}), a low influence on the overall estimation is expected, having fade-out effects among the different azimuth directions.

\begin{figure}[ht]
\begin{center}
\includegraphics[width=0.7\textwidth]{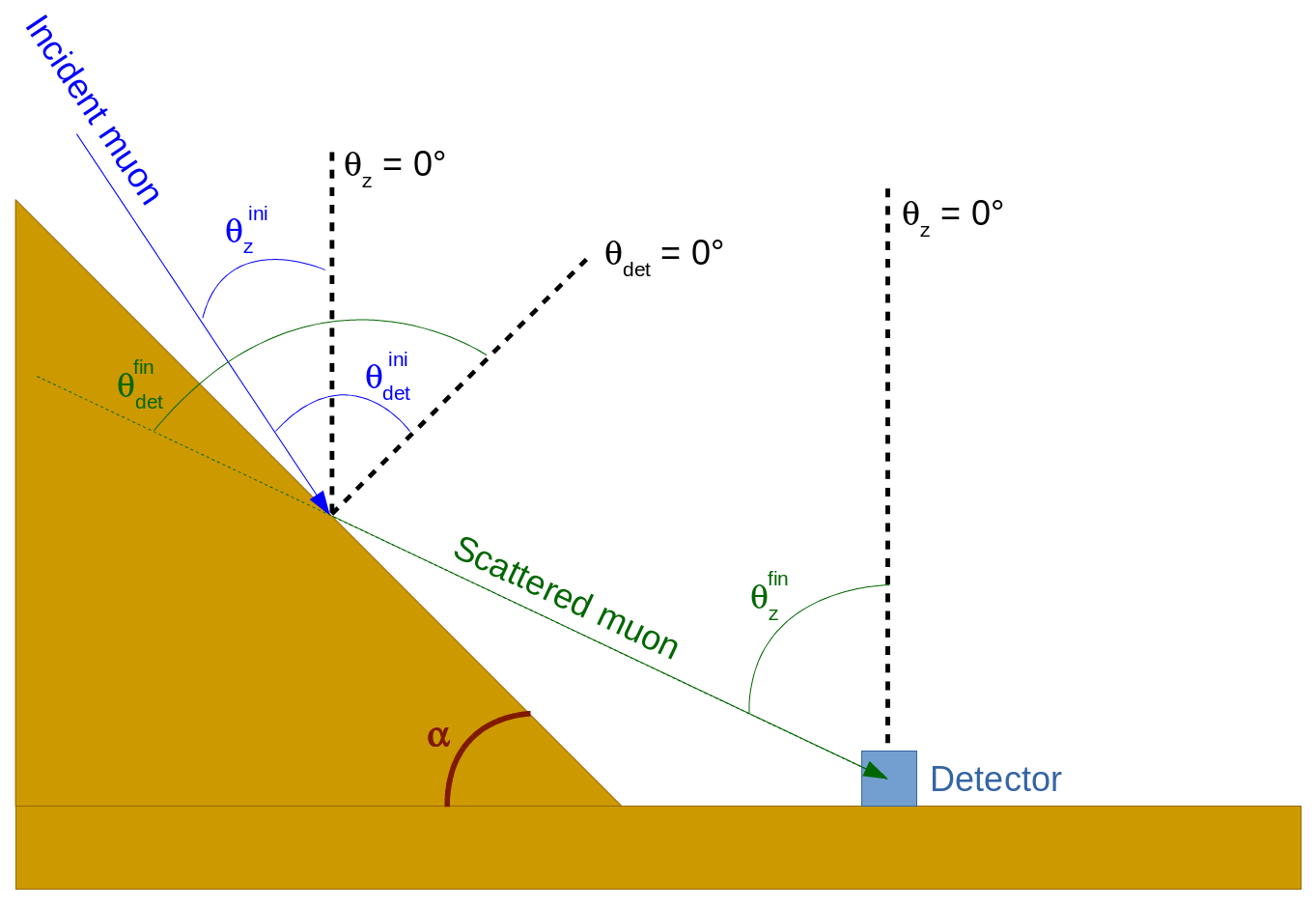}
\caption{Schema showing the relationship between $\theta_{det}$, $\theta_{z}$ and $\alpha$ angles (see text for angles definition), for the use of the muon scattering PDF, $\mathcal{P}$($\theta^{ini}_{det}$, $\theta^{fin}_{det}$, E$^{ini}_{\mu}$), in the B($\theta_{z}$, $\varphi_{z}$) calculation.}
\label{Coordinates_Img}
\end{center}
\end{figure}

With these two assumptions, B($\theta_{z}$, $\varphi_{z}$) corresponds to the product between the initial flux of muons which can be scattered by the corresponding probability to be scattered with a final zenith angle $\theta_{z}$. As already shown, only muons up to 5 GeV with an incident zenith angle higher than 85$^{\circ}$ with respect to the surface normal need to be considered for the forward scattering studies. This delimits the energy and zenith angle ranges to estimate the initial muon flux. The scattering PDF, $\mathcal{P}$($\theta^{ini}_{det}$, $\theta^{fin}_{det}$, E$^{ini}_{\mu}$), corresponds  to the one presented in figure \ref{Diff_PDF_Ewindows_Img}. This PDF was generated using the coordinates $\theta_{det}$ - $\varphi_{det}$, centred in the scattering point and orthogonal to the surface. In order to be able to use this PDF with the $\theta_{z}$ - $\varphi_{z}$ coordinates, it is necessary to define the relationship between $\theta_{det}$ and $\theta_{z}$, which is given by $\theta_{det}$ = $\alpha$ + $\theta_{z}$. $\alpha$ represents the elevation angle of the scattering surface (that means, with respect to the Earth's surface). $\theta_{det}$, $\theta_{z}$ and $\alpha$ angles are presented in figure \ref{Coordinates_Img}. For the case of $\varphi_{z}$ = 0$^{\circ}$, $\alpha$ corresponds to the slope of the lateral surface. For the cases where $\varphi_{z} \neq$ 0$^{\circ}$, it is estimated from the tangent to the tumulus surface at the scattering point. Thus, the expected background B($\theta_{z}$, $\varphi_{z}$) is calculated as:

\begin{equation}
B (\theta_{z}, \varphi_{z}) = \int_{E = 0}^{5} \int_{\theta = 85 - \alpha}^{90 - \alpha} \mathcal{P}(\theta , \alpha + \theta_{z}, E) \phi_{\mu}(E, \theta, \varphi_{z}) dE d\theta
\label{eq:Background}
\end{equation}

For the different muon flux calculations required to obtain S($\theta_{z}$, $\varphi_{z}$) and B($\theta_{z}$, $\varphi_{z}$), the parametrization proposed in \cite{Shukla} has been used, corresponding to:

\begin{eqnarray}
\phi_{\mu}(\theta, E) = I_{0} (n-1) E_{0}^{n-1} (E_{0} + E)^{-n} \left (1 + \frac{E}{\epsilon}\right )^{-1} D({\theta})^{-(n-1)}
\label{eq:Shukla1}
\\
D(\theta) = \sqrt{\frac{R^{2}}{d^{2}} cos^{2}\theta + 2\frac{R}{d} + 1} - \frac{R}{d}cos\theta
\label{eq:Shukla2}
\end{eqnarray}

\noindent where the experimental parameters, summarized in table \ref{Shukla_Coeff} together with other constants used in the equations, have been obtained from the fit of different experimental measurements. This parametrization provides an analytical formula for the muon flux estimation valid for low energy muons and high incident zenith angles.

\begin{table}[]
\centering
\begin{tabular}{cll}
Parameter	&	Value		&				\\
\hline
$E_{0}$		&	4.28$\pm$0.05	& GeV 				\\
$\epsilon$	&	854 		& GeV				\\
$I_{0}$		&	88$\pm$2.4	& m$^{-2}$s$^{-1}$sr$^{-1}$	\\
$n$		&	3.09$\pm$0.03	&				\\
$R/d$		&	174$\pm$12	&				\\
\hline
\end{tabular}
\caption{Values of the parameters and constants used for the estimation of the muon flux based on equations \ref{eq:Shukla1} and \ref{eq:Shukla2}.}
\label{Shukla_Coeff}
\end{table}

\begin{figure}[ht]
\begin{center}
\includegraphics[width=0.9\textwidth]{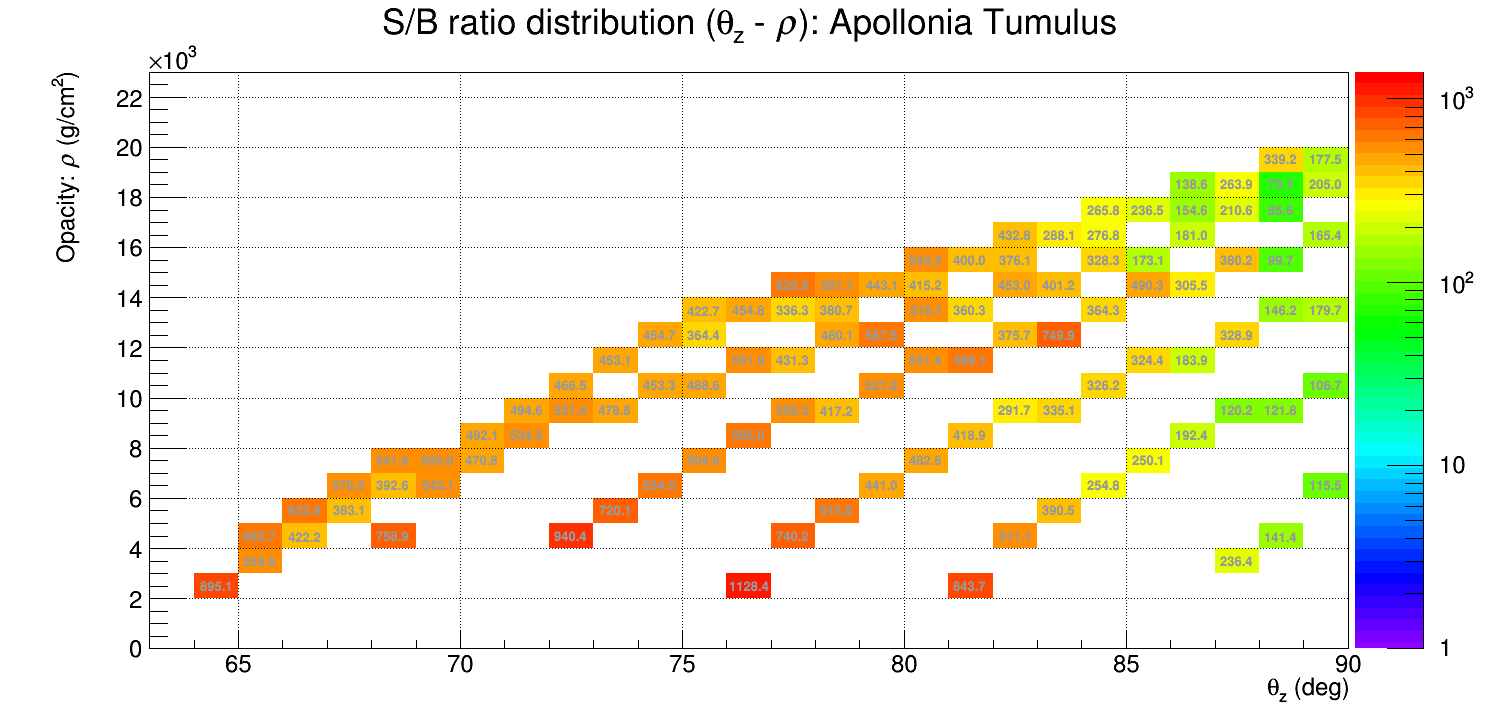}
\caption{Distribution of the ratio between the non-scattered and scattered detected muons (defined as S/B in the text) with respect to the reconstructed zenith incident angle ($\theta_{z}$) and the opacity ($\varrho$) for the Apollonia tumulus case.}
\label{Apollonia_SB_Img}
\end{center}
\end{figure}

With all these ingredients the S/B ratio for the Apollonia tumulus has been calculated scanning the $\varphi_{z}$ range in 10$^{\circ}$ steps and the corresponding $\theta_{z}$ values for each case in 1$^{\circ}$ steps. The results from these calculations are summarized in figure \ref{Apollonia_SB_Img} as a function of $\theta_{z}$ and the opacity $\varrho$, which is a more significant variable than $\varphi_{z}$ since the muon flux is basically independent of the azimuth angle. The main conclusion is that for all the studied directions the S/B ratio is higher than 73.9, which means that at most the 1.3 \% of the detected muons have been previously scattered on the object surface. It is observed that the directions with the lowest S/B values are those with high $\theta_{z}$ values. For these directions lower values for the signal are expected since they correspond to the most horizontal ones (where the muon flux is lower) and, due to the tumulus geometry, these are cases for which longer tumulus length is traversed. Actually, for directions with $\theta_{z}$ lower than 85$^{\circ}$, the S/B ratio is always higher than 254.8, reducing the contribution of the scattered muons to the total detected to less than 0.4 \%. In this region, the obtained S/B values can be considered homogeneous. Differences between directions are basically associated to the uncertainties in the muon scattering PDF. As mentioned in section \ref{sec:Scattering}, even if the used PDF was generated with another target material than the assumed tumulus composition, it would have a limited effect on the results. This leads to consider that the forward scattered muons on the object surface do not significantly influence the results of the muon imaging for the case of tumuli and, by extension, of other objects with similar dimensions and composition.

\subsection{Volcanology: La Soufri\`ere}
\label{sec:Volcanology}

The use of muon imaging for the scanning of volcanoes is another application of this technique, which implies the study of objects with larger dimensions than for archaeology. With this purpose, some projects have already performed measurements in different locations. One of them is the DIAPHANE collaboration \cite{diaphane}, which surveys \textit{La Soufri\`ere} volcano paying special attention to possible variations of the inner liquid/vapour content that can be related to the hydrothermal system dynamics. For this work, \textit{La Soufri\`ere} volcano has been taken as reference to study the impact of the forward scattered muons on the muon imaging reconstruction in volcanology.

As for the case of tumuli, volcanoes geometry can also be approximated to a truncated cone. Based on the topographic plan of \textit{La Soufri\`ere}, their dimensions correspond to a height of 460 m and a base and top radius of 840 m and 160 m respectively. These dimensions lead to a lateral surface with a slope angle of 34.1$^\circ$. In this case a homogeneous composition of standard rock has been considered, with a density of $\rho$ = 2.5 g/cm$^3$, together with 3 different detector positions corresponding to real measurement  points of the DIAPHANE project. They are labelled as h-270, h-170 and h-160 respectively because of the height where they are placed. These positions are summarized in table \ref{Soufriere_DetPos_Tab} taking as reference the centre of the volcano base. Main differences among the positions rely on the distance between the detector and the volcano, going from 5 to 25 m approximately, and on the height with respect to the volcano base, which has a direct influence on the length of the volcano traversed by muons before their detection and, consequently, on the signal S($\theta_z$,$\varphi_z$) computation.
 
\begin{table}[tbp]
\centering
\begin{tabular}{lcccc}
	&	\multicolumn{3}{c}{Detector position}			&	Distance\\
	&	x (m)		&	y (m)		&	z (m)	&	Detector - Volcano (m)\\
\hline
h-270	&	384.86		&	-242.63		&	270	&	14.09	\\
h-170	&	-348.92		&	-482.49		&	170	&	6.72	\\
h-160	&	189.21		&	-598.26		&	160	&	23.99	\\
\hline
\end{tabular}
\caption{Summary of the detector positions with respect to the base centre of \textit{La Soufri\`ere} volcano model (see text for model details).}
\label{Soufriere_DetPos_Tab}
\end{table}

Both tumulus and volcano have been approximated to the same geometrical shape with their corresponding dimensions. So for volcanoes, the procedure to determine the S/B ratio for different incident directions is equivalent to the described in section \ref{sec:Arche} for the tumulus case. The only difference is that for this case the assumed density is $\rho$ = 2.5 g/cm$^3$ instead of $\rho$ = 2.2 g/cm$^3$ (corresponding to the standard soil), affecting in the opacity estimation. Nevertheless, this density variation is expected to have a reduced impact on the results as it has been estimated in section \ref{sec:Scattering}.

\begin{figure}[]
\begin{center}
\includegraphics[width=0.9\textwidth]{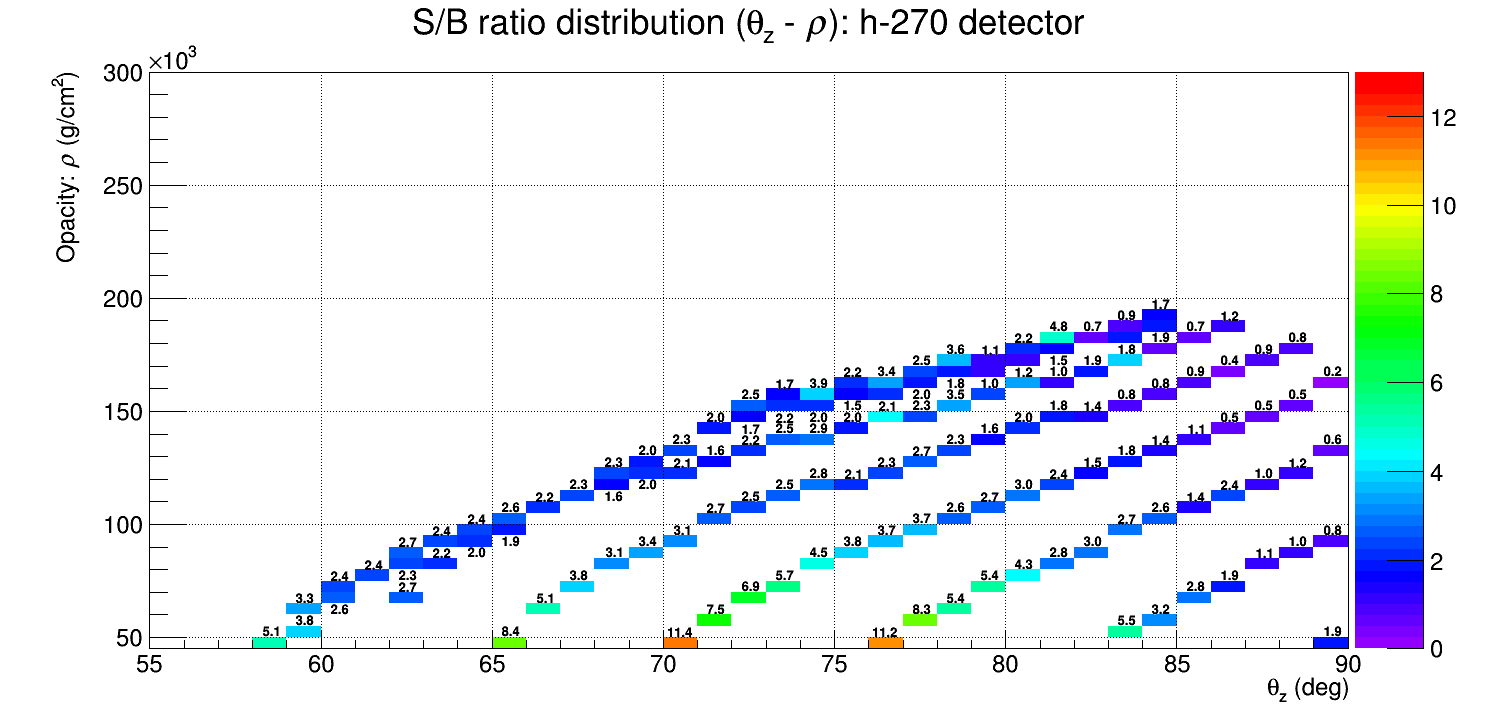}
\includegraphics[width=0.9\textwidth]{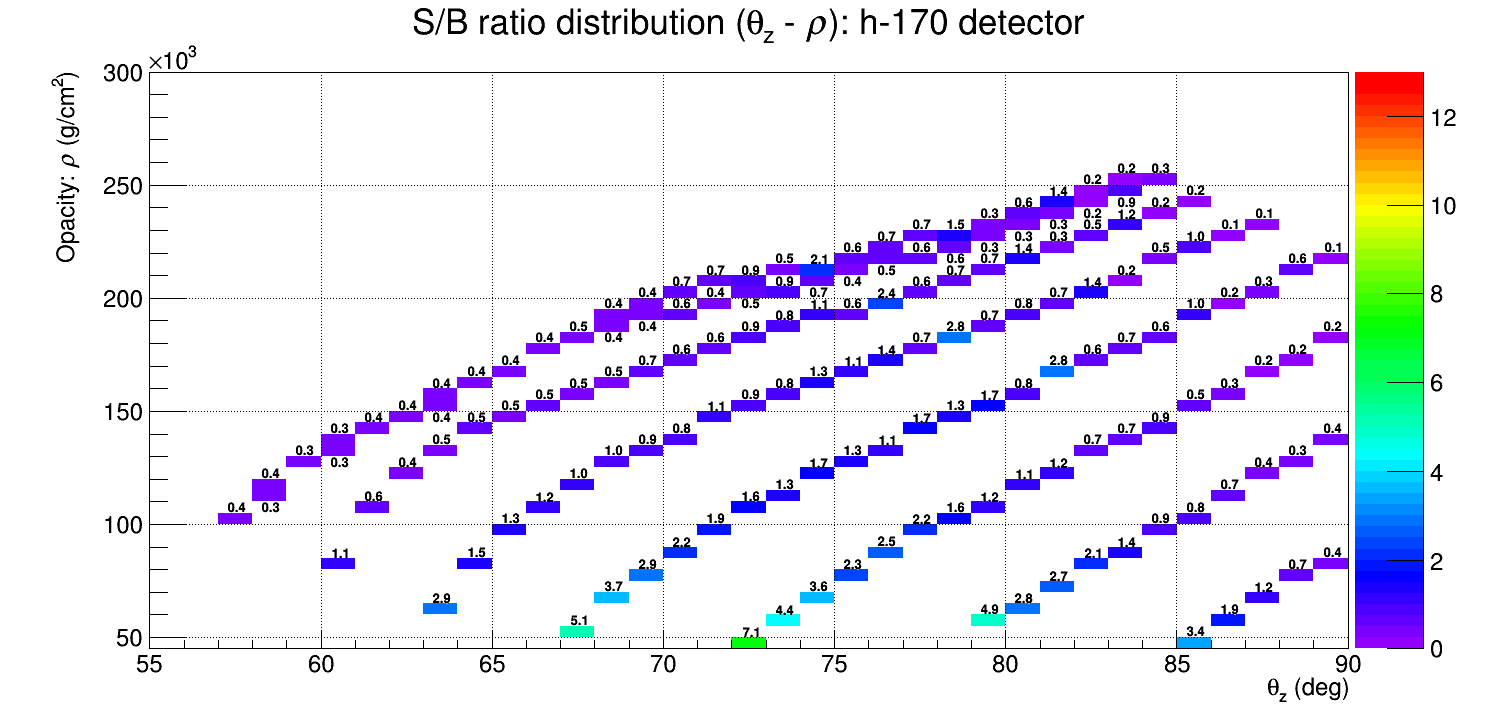}
\includegraphics[width=0.9\textwidth]{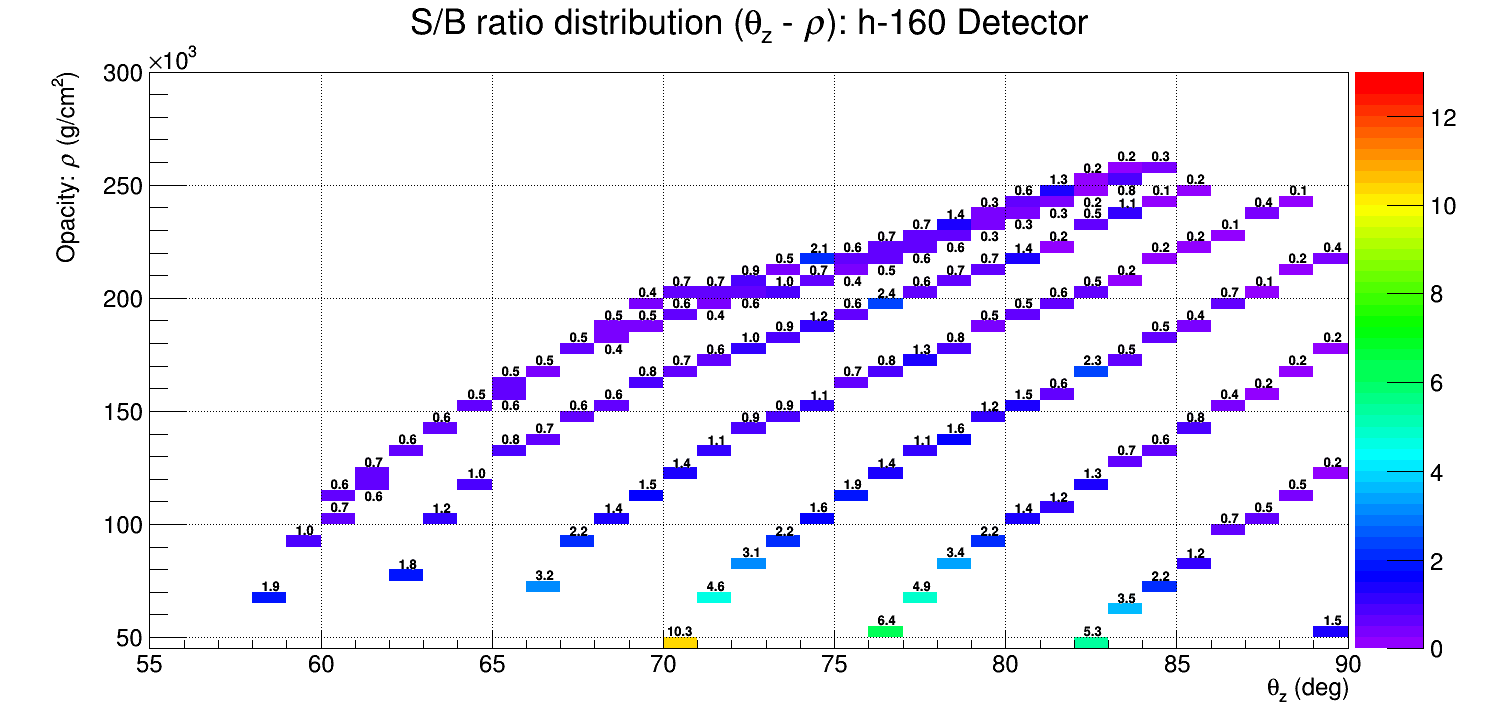}
\caption{Distribution of the ratio between the non-scattered and scattered detected muons (defined as S/B in the text) with respect to the reconstructed zenith incident angle ($\theta_{z}$) and the opacity ($\varrho$). The distribution is showed for the 3 detectors installed in \textit{La Soufri\`ere} volcano. Numbers correspond to the bin value and have been placed next to the corresponding bin to ease their reading.}
\label{Soufriere_SB_Img}
\end{center}
\end{figure}

As for the tumulus case, the S/B ratio have been evaluated scanning the $\varphi_{z}$ range in 10$^{\circ}$ steps and the corresponding $\theta_{z}$ values for each case in 1$^{\circ}$ steps. Results have also been represented with respect to $\theta_{z}$ and the opacity $\varrho$. They have been summarized for the three different detector positions in figure \ref{Soufriere_SB_Img}. For the three cases the S/B ratio takes values significantly lower than for the tumulus, although the corresponding distributions present similar features. For example, the S/B values for directions with $\theta_{z}$ > 85$^\circ$ are again lower than for the rest of the directions. Moreover, for all detector positions, directions with low opacity (corresponding to the volcano contour) present systematically higher values of S/B than those directions pointing to the bulk of the volcano.

Focused on each detector, for the h-270 case, incident directions with high $\theta_{z}$ have S/B ratio values in general below 1, which implies that it is possible to detect more forward scattered muons than those emerging from the volcano, significantly influencing the object density reconstruction. On the opposite side, for the directions where the opacity is smaller than 50$\times$10$^3$ g/cm$^2$, S/B ratio takes values higher than 5, so no more than the 17 \% of the detected muons have been previously scattered. If we consider all the other directions, with $\theta_{z}$ < 85$^\circ$ and $\varrho$ > 50 $\times$10$^3$ g/cm$^2$, the S/B distribution is more homogeneous, having a mean value of 2.7 with a standard deviation of 1.4. That means that in average about 27 \% of the detected muons are low energy forward scattered muons. In this case, the scattered muons have a significant impact on the volcano density reconstruction. Assuming the percentage of scattered muons constant for all the scanned directions, and estimating the uncertainty of this percentage from the standard deviation of the S/B mean value, it would imply that the reconstructed density should be corrected by multiplying it by a factor 1.4$^{+0.4}_{-0.1}$.

Results for the detector positions labelled as h-170 and h-160 are similar between them. This suggests that S/B ratio has more dependence with the detector height with respect to the volcano base (170 m for h-170 and 160 m for h-160) than with the distance between the detector and the volcano (6.72 m for h-170 and 23.99 m for h-160). Since both detectors are placed lower than the h-270 case, the mean volcano length traversed by non-scattered muons is longer in this case. This leads to smaller S/B ratios mainly because lower S($\theta_{z}$, $\varphi_{z}$) values. As mentioned, the features of the distribution are equivalent: smaller S/B values for $\theta_{z}$ > 85$^\circ$ and higher for low opacities. If only muon directions with $\theta_{z}$ < 85$^\circ$ and $\varrho$ > 50 $\times$10$^3$ g/cm$^2$ are considered, a mean S/B value of 1.1 is obtained for both cases with a standard deviation of 0.9 and 1.0 for the h-170 and h-160 detector positions respectively. These values reveal a high influence of the low energy forward scattered muons in the overall detection (almost half of the detected muons have been previously scattered). Keeping the assumption of a constant S/B value for all the considered directions, the density correction factors are 1.9$^{+4.1}_{-0.4}$ for the h-170 position and 1.9$^{+9.1}_{-0.4}$ for the h-160 case.
 
Summarizing, for the case of volcanoes, where the length of material to be traversed by muons is longer than for archaeology, the forward scattering of low energy muons and their further detection has a clear influence on the results of the muon imaging. The three studied scenarios and the defined geometry of the volcano as a truncated cone, reveal that the S/B ratio mainly depends on the length of material traversed by non-scattered muons, those considered for S($\theta_z$,$\varphi_z$) computation. Moreover, for a fixed detector position it can be considered S/B as homogeneous for all the incident directions corresponding to the volcano bulk volume, so a global correction factor for the reconstructed density can be applied. The main source of uncertainty of the S/B ratio estimation comes from those associated to the PDF and consequently, B($\theta_{z}$, $\varphi_{z}$). For this reason, as deduced for the h-270 detection position, a higher S/B mean value translates to a more accurate determination of the correction factor for the reconstructed volcano density.

\section{Summary and discussion}
\label{sec:conclusions}

At present, muon imaging is being used and proposed for an increasing number of different applications. This implies that objects with quite different dimensions can be scanned, from some tens to several hundreds of meters as typical sizes. Furthermore, the composition and density of these objects can also vary from one to other. All the experimental approaches to do muon imaging, generally known as transmission and deviation tomography respectively, rely on the direction reconstruction of the detected muons. For this reason, specially for the transmission-based technique, muons changing their direction because of a scattering on the object surface before their detection, represent an irreducible background noise, biasing the object mean density reconstruction. An estimation of the percentage of these forward scattered muons out of all detected, would allow the estimation of correction factors to reconstruct the proper density.

Muons trajectory deviation is mainly driven by the multiple Coulomb scattering. The resulting angular distribution due to this effect is theoretically described by the Moli\`{e}re theory. Besides, some analytical descriptions of the process have been already performed for particular objects and compositions. Nevertheless, the large variety of objects that are currently proposed to be studied by muon tomography, requires more versatile tools to evaluate the forward scattering muons, easily adaptable to each case. With this aim a set of Monte Carlo simulations have been performed using the Geant4 framework.

These simulations provide a general evaluation of the muon forward scattering probability depending on their energy and their incident angle, being in overall agreement with theoretical estimations. They revealed that muons with energies lower than 5 GeV and incident angles above 85$^\circ$ with respect to the normal direction of the surface, are almost the only muons susceptible to be scattered and then detected. The simulations results have been used as PDF to evaluate the influence of scattered muons in different scenarios. To do that, the signal to background ratio (S/B) has been defined. S($\theta_z$,$\varphi_z$) corresponds to the flux of those muons reconstructed in a direction $\theta_z$-$\varphi_z$  without any previous scattering, while B($\theta_z$,$\varphi_z$) is the muon flux of muons reconstructed with the same direction but after a previous forward scattering on the object surface. The S/B evaluation has been presented for two particular cases, corresponding to two of the applications of muon imaging: archaeology and volcanology.

Taking the muon distribution at Earth's surface proposed in \cite{Shukla}, for the archaeological applications, the Apollonia tumulus has been considered as reference, placing the detector beside the tumulus base. S/B estimations reveals that the percentage of scattered muons detected is never higher than 1.6 \%, being lower than 0.5 \% if the incident zenith angle is smaller than 85$^\circ$. This leads to conclude that the influence of scattered muons for these cases can be neglected.

This is not the case for volcanology applications. A model based on \textit{La Soufri\`ere} volcano, already scanned inside the DIAPHANE project, has been used together with three different detector positions, corresponding to real measurement points. It has been observed a significant influence of the forward scattered muons in the measurement, which can represent up to 50 \% of the detected muons, and even more for incident zenith angles higher than 85$^\circ$. S/B values can be considered homogeneous for the directions corresponding to the bulk volume of the volcano. Main differences on S/B mainly depend on the height of the detector with respect to the volcano base. Due to the volcano geometry, defined as a truncated cone, this is directly related to the muon path length along the volcano. Other features, such as the distance between the detector and the volcano, seem to have smaller influence. With the estimations and numbers obtained in this work, correction factors for the density reconstruction have been computed, taking values from 1.4 to 1.9 depending on the detector position.

Forward scattered muons represent events that are in principle not taken into account, so their detection has a direct impact on the mean density reconstruction. Nonetheless, the observed homogeneity on the S/B ratio for all the considered directions, both in the tumulus and volcano case, leads to think that these muons would not significantly affect fading the resolution of the resulting image.

All these estimations and conclusions are based on simulations of scattered muons on standard rock, which has been demonstrated to produce equivalent results than the standard soil case. In any case, changing accordingly the material composition and properties, this simulation framework can be used to evaluate the influence of forward scattered muons for further muon imaging measurements of other objects and structures.

\acknowledgments

Authors would like to acknowledge the financial support from the UnivEarthS Labex program of Sorbonne Paris Cit\'{e} (ANR-10-LABX-0023 and ANR-11-IDEX-0005-02). Data from \textit{La Soufri\`ere} volcano are part of the ANR DIAPHANE project ANR-14-ce04-0001. Part of the project has been funded by the INSU/IN2P3 TelluS and ``DEFI Instrumentation aux limites'' programmes.


\begin{thebibliography}{99}

\bibitem{Anderson1} S.H. Neddermeyer, C.D.Anderson. \emph{Note on the nature of cosmic-ray particles}, \emph{Phys. Rev.} \textbf{51} (1937) 884

\bibitem{Anderson2} S.H. Neddermeyer, C.D.Anderson.  \emph{Cosmic-ray particles of inter-mediate mass}, \emph{Phys. Rev.} \textbf{54} (1938) 88

\bibitem{Auger} P. Auger. \emph{Les rayons cosmiques}, \emph{PUF, Paris} (1941) 136

\bibitem{Nagamine} K. Nagamine. \emph{Introductory Muon Science}, \emph{Cambridge University Press, Cambridge} (2003)

\bibitem{Lesparre_2010} N. Lesparre et al. \emph{Geophysical muon imaging: feasibility and limits}, \emph{Geophysical Journal International} \textbf{183} (2010) 1348

\bibitem{Marteau_2012} J. Marteau et al. \emph{Muons tomography applied to geosciences, volcanology} \emph{Nucl. Instrum. Meth. A} \textbf{695} (2012) 23

\bibitem{Borozdin} K.N. Borozdin et al. \emph{Radiographic imaging with cosmic-ray muons} \emph{Nature} \textbf{422} (203) 277

\bibitem{Procureur} S. Procureur. \emph{Muon imaging: Principles, technologies and applications} \emph{Nucl. Instrum. Meth. A}
\url{http://dx.doi.org/10.1016/j.nima.2017.08.004}

\bibitem{George} E.P. George. \emph{Cosmic rays measure overburden of tunnel}, \emph{Commonwealth Engineer} \textbf{455} (1955)

\bibitem{Alvarez} L.W. Alvarez. \emph{Search for hidden chambers in the pyramids using cosmic rays}, \emph{Science} \textbf{167} (1970) 832

\bibitem{SPNature} K. Morishima et al. \emph{Discovery of a big void in Khufu's Pyramid by observation of cosmic-ray muons}, \emph{Nature}
\url{http://dx.doi.org/10.1038/nature24647} (2017)

\bibitem{Malmqvist} L. Malmqvist et al. \emph{Theoretical studies of in-situ rock density determination using cosmic-ray muon intensity measurements with application in mining geophysics}, \emph{Geophysics} \textbf{44} (1979) 1549

\bibitem{Kume} N. Kume et al. \emph{Muon trackers for imaging a nuclear reactor}, \emph{J of Instr.} \textbf{11} (2016) P09008

\bibitem{nishiyama} R. Nishiyama et al. \emph{Monte Carlo simulation for background study of geophysical inspection with cosmic-ray muons}, \emph{ Geophys. J. Int.} \textbf{206} (2016) 1039

\bibitem{3d_souf} M. Rosas-Carbajal et al. \emph{Three-dimensional density structure of La Soufri{\`e}re de Guadeloupe lava dome from simultaneous muon radiographies and gravity data}, \emph{Geophysical Research Letters} \textbf{44} (2017) 6743

\bibitem{Bethe} H.A. Bethe. \emph{Moli\`{e}re's Theory of Multiple Scattering}, \emph{Physical Review} \textbf{89} (1953) 1256

\bibitem{Schneider} U. Schenider et al. \emph{Coulomb scattering and spatial resolution in proton radiography}, \emph{Medical Physics} \textbf{21} (1994) 1657

\bibitem{Geant4} S. Agostinelli et al. (GEANT4 collaboration), \emph{GEANT4: A Simulation toolkit}, \emph{Nucl. Instrum. Meth. A} \textbf{506} (2003) 250, \emph{IEEE Trans. Nucl. Sci.} \textbf{53} (2006) 270\\
J. Allison et al. \emph{Recent developments in GEANT4}, \emph{Nucl. Instrum. Meth. A} \textbf{835} (2016) 186

\bibitem{PDG} C. Patrignani et al. (Particle Data Group), \emph{Review on Particle Physics}, \emph{Chin. Phys. C} \textbf{40} (2016) 100001

\bibitem{gomez2016arche} H. G\'{o}mez et al. \emph{Studies on muon tomography for archaeological internal structures scanning}, \emph{Journal of Physics: Conference Series}, \textbf{718(5)} (2016) 052016

\bibitem{jourde2013experimental} K. Jourde et al. \emph{Experimental detection of upward going cosmic particles and consequences for correction of density radiography of volcanoes}, {\em Geophysical Research Letters} \textbf{40(24)} (2013) 6334

\bibitem{jourde_nature_2016} K. Jourde et al. \emph{Muon dynamic radiography of density changes induced by hydrothermal activity at the La Soufri\`{e}re of Guadeloupe volcano}, {\em Scientific Reports} \textbf{6} (2016)  33406

\bibitem{ScanPyramids} ScanPyramids project, \url{www.scanpyramids.org}

\bibitem{Arche} H. G\'omez et al. \emph{Muon imaging for archaeological applications: feasibility studies of the Apollonia Macedonian Tumulus}, \emph{(In preparation)}

\bibitem{PDG_muons} D.E. Groom et al. \emph{Muon stopping-power and range tables}, \emph{Atomic Data and Nuclear Data Tables} \textbf{78} (2001) 183

\bibitem{Shukla} P. Shukla. \emph{Energy and angular distributions of atmospheric muons at the Earth}, arXiv:1606.06907 [hep-ph]

\bibitem{diaphane} Diaphane project, \url{www.diaphane-muons.com}

\end{thebibliography}
\end{document}